\documentclass[12pt]{article}
\usepackage{a4wide}
\usepackage{amssymb}
\usepackage{graphicx}
\def\makeatletter{\catcode`\@=11}
\makeatletter
\def\mathbox#1{\hbox{$\m@th#1$}}%
\def\math@ccstyles#1#2#3#4#5#6#7{{\leavevmode
      \setbox0\mathbox{#6#7}%
      \setbox2\mathbox{#4#5}%
      \dimen@ #3%
      \baselineskip\z@\lineskiplimit#1\lineskip\z@
      \vbox{\ialign{##\crcr
             \hfil \kern #2\box2 \hfil\crcr
             \noalign{\kern\dimen@}%
             \hfil\box0\hfil\crcr}}}}
\def\mathaccstyles{\math@ccstyles\maxdimen}
\def\maththroughstyles{\math@ccstyles{-\maxdimen}}
\def\unity%
 {\maththroughstyles{.45\ht0}\z@\displaystyle {\mathchar"006C}\displaystyle 1}
\begin{document}

\rightline{QMUL-PH-09-24}
\vspace{2truecm}

\centerline{\LARGE \bf M5 spikes and operators}
\vspace{.7truecm}
\centerline{\LARGE \bf  in the HVZ membrane theory} 
\vspace{1.3truecm}

\centerline{
    {\large \bf D. Rodriguez-Gomez}
                                                       }

\vspace{.4cm}
\centerline{{\it Queen Mary, U. of London}}
\centerline{{\it Mile End Road}}
\centerline{{\it London, E1 4NS}}
\centerline{{\it U.K.}}
\vspace{0.2cm}
\centerline{{\tt D.Rodriguez-Gomez@qmul.ac.uk}}
\vspace{2truecm}
                     
\centerline{\bf ABSTRACT}
\vspace{.5truecm}
In this note we study some aspects of the so-called dual ABJM theory introduced by Hanany, Vegh \& Zaffaroni. We analyze the spectrum of chiral operators, and compare it with the spectrum of functions on the mesonic moduli space $\mathcal{M}=\mathbb{C}^2\times \mathbb{C}^2/\mathbb{Z}_k$, finding expected agreement for the coherent branch. A somewhat mysterious extra branch of dimension $N^2$ opens up at the orbifold fixed point. We also study BPS solutions which represent M2/M5 intersections. The mesonic moduli space suggests that there should be two versions of this spike: one where the M5 lives in the orbifolded $\mathbb{C}^2$ and another where it lives in the unorbifolded one. While expectedly the first class turns out to be like the ABJM spike, the latter class looks like a collection of stacks of M5 branes with fuzzy $S^3$ profiles. This shows hints of the appearance of the global $SO(4)$ at the non-abelian level which is otherwise not present in the bosonic potential. We also study the matching of SUGRA modes with operators in the coherent branch of the moduli space. As a byproduct, we present some formulae for the laplacian in conical $CY_4$ of the form $\mathbb{C}^n\times CY_{4-n}$.

\noindent

\newpage

\section{Introduction}

Big progress has been made towards understanding three-dimensional SCFTs and their gravitational $AdS_4$ duals. The pioneering works of Bagger and Lambert \cite{Bagger:2007jr} and Gustavsson \cite{Gustavsson:2007vu} culminated with the celebrated theory discovered by Aharony, Bergman, Jafferis and Maldacena (henceforth ABJM) \cite{Aharony:2008ug}. The key ingredient to these constructions is a Chern-Simons kinetic term for the gauge fields. This had been anticipated by Schwarz in \cite{Schwarz:2004yj} on the basis that, while the 3d Yang-Mills coupling is dimensionful, a classically marginal Chern-Simons kinetic term can be constructed, thus being the natural candidate for a conformal theory. Ever since, a lot of progress has been made towards a better understanding of the $AdS_4/CFT_3$ duality.

Following ABJM, the 3d SCFT's dual to $AdS_4$ backgrounds are Chern-Simons-matter theories. Generically, they have $g$ gauge groups of equal rank\footnote{Unequal ranks have been also discussed (\textit{e.g.} \cite{Aharony:2008gk}). However, we focus on the minimal case where no fractional branes are present, such that all the ranks have equal (and large) rank.}, each one equipped with a CS kinetic term whose CS level is $k_i$. In order to have a four-dimensional complex moduli space, we should require the sum of the $k_i$ to vanish. It turns out that the role of the CS level is far from trivial. In particular, it imposes discrete identifications leading to an orbifold-type moduli space; and it is only for the smallest possible value of  the $k$'s that the moduli space is the unorbifolded version. For example, in the ABJM case, the moduli space is $\mathbb{C}^4/\mathbb{Z}_k$, where the orbifold acts ``diagonal" on the $\mathbb{C}^4$. For generic $k$ this orbifold preserves $\mathcal{N}=6$ SUSY. It is only for $k=1,2$ that the SUSY is enhanced to $\mathcal{N}=8$, as so does the R-symmetry to $SO(8)$. This symmetry enhancement is a quantum-mechanical phenomenon due to monopole operators which can be used to construct chiral primaries furnishing $SO(8)$ representations. The lesson is then that the explicit symmetry of the theory is that of the generic $k$ case. Another interesting example is the theory dual to M2 branes probing $\mathcal{C}(Q^{111})$. While the global symmetry of that $CY_4$ is $SU(2)^3$, the proposed dual theory \cite{Franco:2008um, Franco:2009sp} only exhibits an explicit $SU(2)$. \footnote{There have been other proposals in the literature, such as \cite{Aganagic:2009zk}. A classification of quivers containing a survey of the possibilities can be found in \cite{Davey:2009sr}. Very recently other candidates involving fundamental matter have been proposed in \cite{Jafferis:2009th, Benini:2009qs}.} The reason is that the theory, for generic CS levels, describes an orbifold of $\mathcal{C}(Q^{111})$ which does only preserve precisely one $SU(2)$. In retrospective, this lack of explicit symmetry was crucial with respect to previous attempts \cite{Fabbri:1999hw} to find the dual theory to M2 branes moving on $\mathcal{C}(Q^{111})$. Similar comments hold for $M^{3,2}$ \cite{Martelli:2008si,Franco:2009sp}.

Soon after the ABJM theory was discovered, another candidate describing, at $k=1$, branes in flat space was found by Hanany, Vegh and Zaffaroni (henceforth HVZ) \cite{Hanany:2008fj}. At generic CS levels $k$, this theory has a mesonic moduli space $\mathbb{C}^2\times \frac{\mathbb{C}^2}{\mathbb{Z}_k}$; which should preserve $\mathcal{N}=4$ SUSY. At the same time it has a classically marginal superpotential. Since, as described in \cite{Gaiotto:2007qi}, $\mathcal{N}\ge 3$ SUSY requires non-trivial relations between the superpotential coupling constant and the (trivially renormalized) CS level, this would automatically imply that the HVZ theory is conformal at the quantum level. This is a strong statement, since in 3d, where there is no analogue of $a$-maximization\footnote{The exact superconformal R-charge should minimize the $\tau$ coefficients in 2-point functions of global currents \cite{Barnes:2005bm}. However, lack of powerful tools such as anomaly cancellation has prevented to formulate a useful version of $\tau$ maximization similar to \cite{Intriligator:2003jj}.}, it is quite non-trivial to ensure the existence of a conformal fixed point in a theory with SUSY less than $\mathcal{N}=3$.

In \cite{Aharony:2008ug} and \cite{Benna:2008zy} it was shown how the component lagrangian of the ABJM theory re-arranges itself such that a global $SU(4)\sim SO(6)$ symmetry, not naively present in the $\mathcal{N}=2$ superspace formulation, actually appears. Since this symmetry does not commute with SUSY, it is an R-symmetry. Therefore, being $SO(6)$ the R-symmetry, this automatically implies that the ABJM theory does indeed preserve $\mathcal{N}=6$ SUSY. Following that example, one would expect that the HVZ theory, written in components, enjoys an $SO(4)$ R-symmetry corresponding to the desired $\mathcal{N}=4$ SUSY. However, it turns out that such $SO(4)$ is not present in the bosonic potential of the theory. Another (most possibly related) issue was found in \cite{Choi:2008za}, where the superconformal index for the HVZ theory was computed at large $k$ and shown to match almost but not quite the SUGRA expectations. This raises some puzzles which we try to address in this note.

One particular tool which we will use are some BPS solutions. One of the seeds for the ``M2 minirevolution" was the M2-M5 intersection studied by Harvey and Basu \cite{Basu:2004ed}. First in \cite{Terashima:2008sy} and subsequently in \cite{Gomis:2008vc}, this intersection was analyzed under the light of ABJM. Further developments in \cite{Hanaki:2008cu} and \cite{Nastase:2009ny} showed that the M5-spike, which has a transverse $S^3$ profile, wraps the $S^1$ orbifolded by the large $k$ limit. The $S^3$ appears therefore as a Hopf fibration of a circle over an $S^2$, such that, in the large $k$ limit, the IIA picture is a D2-D4 intersection \cite{Nastase:2009ny}. In the HVZ case, given the form of the abelian moduli space, the situation should be richer. On one hand, we expect the possibility of having M5 expanding in the $\mathbb{C}^2/\mathbb{Z}_k$ part of the geometry which should lead to an $S^3$ as a Hopf fibration in very much of the spirit as in ABJM. On the other hand, there should be the possibility of M5's growing in the $\mathbb{C}^2$ piece, which does not feel the $\mathbb{Z}_k$ orbifold. This has the interesting consequence that this intersection should reduce to a D2-NS5 intersection, that is, the M5 should be \textit{transverse} to the M-theory circle. In turn, the matching of these structures with the BPS solutions of HVZ would reassure the fact that it actually describes the desired orbifold; in particular showing hints of the appearance of the global $SO(4)$ at the non-abelian level.

The organization of this note is as follows: in section 2 we review the HVZ theory and study explicitly its chiral ring. In section 3 we turn to study some of its BPS solutions which should indeed represent the M2-M5 intersections described above. In section 4 we study the matching of chiral fields with SUGRA modes, focusing on the short graviton multiplet. We also collect formulae which can be useful to study $CY_4$ conical backgrounds of the form $\mathbb{C}^n\times CY_{4-n}$. We finish with some concluding remarks in section 5.

\section{The dual ABJM}

The dual ABJM theory introduced in \cite{Hanany:2008fj} has the quiver

\begin{center}
\includegraphics[scale=.3]{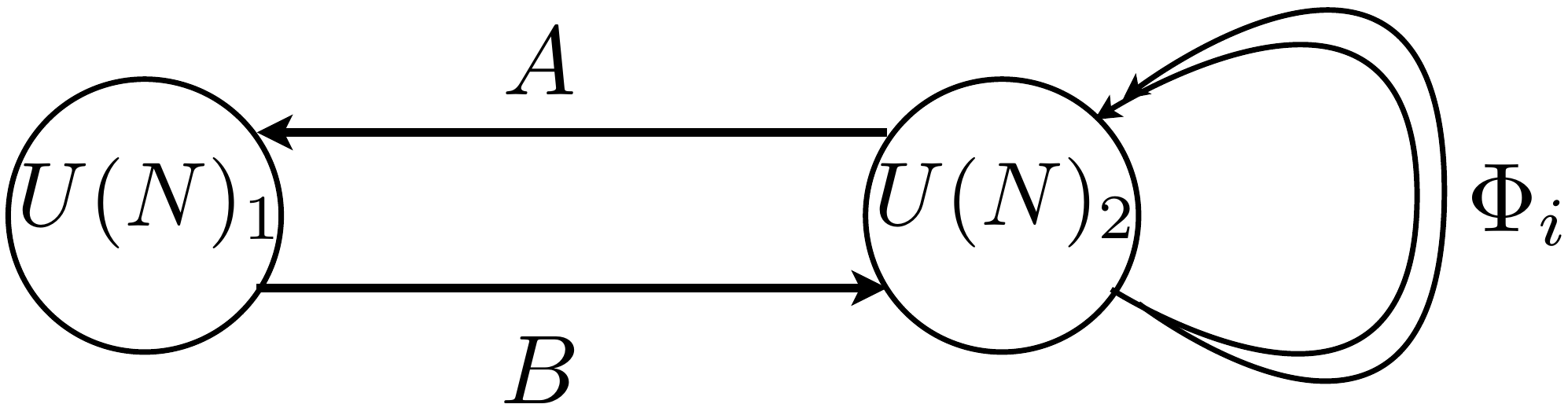}
\end{center}

The superpotential is

\begin{equation}
W={\rm Tr}\Big(\, [\Phi_1,\Phi_2]\, A\, B\,\Big)
\end{equation}
The CS levels are $(-k,k)$, such that they add to zero. We note that being quartic, the superpotential is classically marginal, and thus the theory is classically conformal.

There is a global $SU(2)$ symmetry acting on $(\Phi_1,\,\Phi_2)$. This implies that both $\Phi_1$ and $\Phi_2$ have the same R-charge. 

For future convenience, let us review a few details about $\mathcal{N}=2$ 3d theories. On general grounds, the bosonic potential reads\footnote{The conventions used here differ in the normalization of the CS term as compared with \cite{Aharony:2008ug} or \cite{Benna:2008zy} in that the $k$ we are using contains an extra factor of $(8\pi)^{-1}$, that is $k_{here}=k_{ABJM}/8\pi$. This will be relevant only in section 4.} 

\begin{equation}
\label{V}
V_b={\rm Tr}\Big(-4\sum_gk_g\sigma_gD_g+\sum_gD_g\mu_g-\sum_{X_{ab}}(\sigma_aX_{ab}-X_{ab}\sigma_b)(\sigma_aX_{ab}-X_{ab}\sigma_b)^{\dagger}-\sum_{X_{ab}}|\partial_{X_{ab}}W|^2\Big)
\end{equation}
where the sum over $g$ runs over gauge groups and the sum over $X_{ab}$ runs over all scalars going from node $a$ to node $b$. Furthermore, $\mu_g$ are the usual moment maps of each group, which in this case read

\begin{equation}
\label{momap}
\mu_1=B\, B^{\dagger}-A^{\dagger}\, A\ ,\qquad \mu_2=A\, A^{\dagger}-B^{\dagger}\, B+[\Phi_1,\,\Phi_1^{\dagger}]+[\Phi_2,\,\Phi_2^{\dagger}]
\end{equation}
As it is well-known, both the $D_g$ and the $\sigma_g$ are auxiliary fields which must be integrated out, thus rendering the bosonic potential a sum of squares which is manifestly positive definite. Integrating out the $D_g$ sets

\begin{equation}
\mu_g=4\,k_g\, \sigma_g
\end{equation}
Following by now standard techniques \cite{Jafferis:2008qz, Martelli:2008si, Hanany:2008cd}, when looking at the mesonic moduli space, since all gauge groups are of equal rank, we take our fields, which are square $N\times N$ matrices, to be diagonal, which generically sets $\sigma_g=\sigma\, \forall g$. \footnote{The diagonalization procedure is actually more involved as discussed in \cite{Berenstein:2002ge} and generalized to the 3d case in \cite{Berenstein:2009ay}.} We stress that this not necessarily exhausts the moduli space, since other branches can, and in this case indeed will, exist. 

Thus, the geometric branch moduli space we are interested in is the set of configurations satisfying the F-flatness conditions subject to the restriction that

\begin{equation}
\mu_g=4\, k_g\,\sigma\qquad \forall g
\end{equation}
In the case at hand, since we have just two nodes, it happens that $\mu_1=-\mu_2$. Since $k_1=-k_2$, one of the two equations above is actually redundant; while the other is trivially satisfied, since it just defines the auxiliary field $\sigma$. In addition, once we focus on diagonal fields, it is clear that the superpotential vanishes. Therefore, the moduli space consists of 4 complex fields with no other relation.

However, we still need to take into account the action of the CS level \cite{Martelli:2008si}. In this case, it turns out that it imposes the following identifications

\begin{equation}
(A,B)\sim (e^{i\frac{2\pi}{k}}\, A, e^{-i\frac{2\pi}{k}}\, B)
\end{equation}
Thus, the mesonic moduli space is

\begin{equation}
\mathcal{M}=\mathbb{C}^2\times \frac{\mathbb{C}^2}{\mathbb{Z}_k}
\end{equation}
where the $\mathbb{C}^2$ is parametrized by the adjoints (which do not pick any phase) and the $ \frac{\mathbb{C}^2}{\mathbb{Z}_k}$ is parametrized by $(A,\,B)$.

In view of this mesonic moduli space, one would expect that the global symmetry of HVZ is actually $SO(4)\times SU(2)$, being the $SO(4)$ rotating the $\mathbb{C}^2\sim\mathbb{R}^4$ the R-symmetry. At $k=1$ the $\mathbb{C}^4$ moduli space would suggest an enhancement to an $SO(8)$ global (in fact R) symmetry, which would require all the fields to have $1/2$ R-charge. At higher $k$, one $\mathbb{C}^2$, that parametrized by the adjoints, remains unorbifolded. This $\mathbb{C}^2\sim \mathbb{R}^4$ should be rotated by the $SO(4)_R$. In our $\mathcal{N}=2$ formalism notation only an $SU(2)$ subgroup rotating the $\Phi_i$ is manifest, which is enough to ensure that these two fields have the same R-charge. Thus, since at $k=1$ the moduli space suggests that all fields have R-charge $1/2$, and the higher $k$ orbifold does not act on the $SU(2)$ $\Phi_i$ doublet, it is natural to conjecture that the $\Phi_i$ have $1/2$ R-charge. Another hint in support of this guess is that, since at the abelian level the superpotential vanishes, all the fields appear in a symmetric way, thus suggesting equal R-charge $1/2$ for all of them.

\subsection{Chiral operators}

The $AdS/CFT$ correspondence requires the spectrum of chiral operators in the gauge theory to match the KK harmonics in the geometry side. As it happens with the ABJM theory, we expect that an important role in such a matching is played by monopole operators. The study of such operators is well beyond the scope of this paper, and we will content ourselves with taking the large $k$ limit, where they will not contribute. Under that assumption, the operators we are interested in are gauge-invariants in the usual sense, \textit{i.e.} closed loops in the quiver, modulo F-terms. These read 

\begin{equation}
\label{F}
F_A=B\, [\Phi_1,\Phi_2]=0\ , \quad F_B=[\Phi_1,\Phi_2]\, A=0\ ,\quad F_{\Phi_1}=[\Phi_2,AB]=0\ ,\quad F_{\Phi_2}=[AB,\Phi_1]=0
\end{equation}

It is clear that the fields $A$ and $B$ can only occur in any gauge-invariant operator through the invariant combination $A\,B$. Calling $A\,B=M$, we have that the most generic chiral operator looks like

\begin{equation}
\label{operators}
\mathcal{O}_{i_1\cdots\, i_j|\,q}={\rm Tr}\, \Phi_{i_1}\,\cdots \Phi_{i_j}\, M^q
\end{equation}
Note that the F-term condition $[\Phi_i,\,M]=0$ allows to group together powers of $M$ as in (\ref{operators}). With the R-charge assignation proposed in the previous subsection, we would have that these operators have R-charge $q+\frac{j}{2}$.

We now need to discuss the symmetry properties of the $\{i_q\}$ indices. Let us consider first the case $q\ne 0$. The F-terms imply that $M\, \Phi_i\,\Phi_j=M\, \Phi_j\,\Phi_i$ and $\Phi_i\,\Phi_j\, M=\Phi_j\,\Phi_i\, M$. Thus, without having to refer to $M$ vanishing or not, the indices of the nearest two neighbors to $M^q$ are automatically symmetrized. However, since $M$ commutes with $\Phi_i$, we can bring to that position any two fields. Thus, we have that the latin indices are symmetrized. 

The case $q=0$, not involving $M$, is more subtle. For generic $M$, from the equation $M\,[\Phi_i,\,\Phi_j]=0$, we conclude that $[\Phi_i,\,\Phi_j]=0$; thus ensuring the symmetry of the $\{i_q\}$ indices. However, at the origin of $M$, that is, if $M$ has some zero block, things are more subtle. Let us gauge-fix $M$ to be diagonal. For generic eigenvalues $m_i$ of $M$, as described, the F-terms allow to symmetrize latin indices. However, if some, say $n\le N$, of the $m_i$'s  vanish, we get $n$ directions where $[\Phi_i,\,\Phi_j]$ needs not to vanish. More explicitly

\begin{equation}
M=\left(\begin{array}{ c c c c} m_1 & & &\\ & \ddots & & \\ & & m_{N-n} &\\ & & & 0\times \unity_n\end{array}\right)\, \Rightarrow \,
\Phi_i=\left(\begin{array}{c c c} \phi_i^1 & &\\ & \ddots & \\ & & \phi_i^N\end{array}\right) + \left(\begin{array}{ c c} 0\times \unity_{N-n} & \\ &U_i\end{array}\right)
\end{equation}
being $U_i$ a generically non-vanishing matrix in the complement of the Cartan subalgebra of $SU(n)$. If $U_i\ne 0$, since $[\Phi_i,\,\Phi_j]\ne 0$, latin indices would not be symmetrized; and therefore we would get extra operators in the full chiral ring as compared to the symmetrized case ($U_i=0$). Note that the dimension of this codimension 2 extra branch we would get is of order $N^2$. These states, living in the orbifold fixed point, might be the ones responsible for the index computation mismatch \cite{Choi:2008za}. Indeed a related problem is that, following observations in \cite{Berenstein:2002ge, Berenstein:2009ay}, the structure of the HVZ moduli space does not fit into the structure expected for branes probing a singularity. \footnote{We thank D.Berenstein for explanations on this point.} From now on we will focus on the ring of operators satisfying the F-terms for generic field values, that is $U_i=0$ such that $[\Phi_i,\,\Phi_j]=0$. This ring is sometimes called the coherent part of the moduli space \cite{Forcella:2008bb}. We could think of it as composed by those operators of the form (\ref{operators}), such that we can symmetrize latin indices and afterwards take $q=0$.

The chiral ring of operators should match the KK harmonics of $\mathcal{M}=\mathbb{C}^2\times \mathbb{C}^2/\mathbb{Z}_k$. This orbifold is defined as

\begin{equation}
\mathcal{M}=\{ (z_1,z_2,z_3,z_4)\in \mathbb{C}^4\,/\,(z_1,z_2,z_3,z_4)\sim (z_1,z_2,e^{i\frac{2\pi}{k}}z_3, e^{-i\frac{2\pi}{k}}z_4)\}
\end{equation}
The equivalence relation of the orbifold forces $z_3\, z_4$ to appear together. Let us call that monomial $m=z_3z_4$. Then, it is clear that the functions on $\mathcal{M}$ are

\begin{equation}
f_{i_1\,\cdots\, i_n|\, q}=z_{i_1}\, \cdots\, z_{i_n}\, m^q
\end{equation}
which is clearly in one to one correspondence with (\ref{operators}). 

Note that in the large $k$ limit we lose one dimension and indeed the coordinate ring loosely speaking looks like that of $\mathbb{C}^3$. We can also see this from the Hilbert series point of view. \footnote{We are grateful to Amihay Hanany for conversations on this point.} This object encodes the number of chiral operators in the coherent branch, which coincides with the number of holomorphic functions on the variety. For an introduction in the context of pure algebraic geometry, the reader might consult $e.g.$ \cite{AlgebraicGeometry}. In the context of gauge theories, the Hilbert series is a well-studied object (see \textit{e.g} \cite{Benvenuti:2006qr, Feng:2007ur, Forcella:2008bb}); which has also been considered in the context of 3d SCFTs (see $e.g.$ \cite{Hanany:2008fj, Hanany:2008cd}).

The Hilbert series depends on a parameter $t$, which is to be thought as $t=e^{-\mu}$, such that $\mu$ admits an interpretation as a (not necessarily normalized) chemical potential for the R-charge. Thus, the coefficients of a given power of $t$ in the expansion of the Hilbert series count the number of operators with given R-charge. In the situation at hand, considering the $U(1)$ case, one can convince oneself that the Hilbert series for the HVZ theory is 

\begin{equation}
g=\frac{(1-t^{2\,k})}{(1-t^2)\, (1-t^k)^2\, (1-t)^2}
\end{equation}

At the $U(1)$ level one can verify that the number of gauge-invariant operators at generic $k$ with respect to the effective D-terms $\mu_g=4\,k_g\,\sigma$ matches the coefficients in the expansion of $g$. In order to give a flavor, let us consider the simplest cases $k=1,\, 2$:

\begin{equation}
g_{k=1}=1+4\, t+ 10\, t^2+\cdots\ ; \qquad g_{k=2}=1+2\, t+6\, t^2+\cdots
\end{equation}

On the other hand, since $\sum k_g=0$, out of the $\mu_g=4\,k_g\,\sigma$ equations only one of them is independent. However, that equation is just the definition of the auxiliary field $\sigma$, and as such is always satisfied. Thus, at the abelian level, the relevant operators to consider are made out of the fields with no other restriction apart from the action of the higher $k$ orbifold. Thus, at $k=1$ we would have the following operators

\begin{equation}
\{ \{1\},\, \{\Phi_1,\,\Phi_2,\, A,\, B\},\,\{\Phi_1^2,\, \Phi_2^2,\, \Phi_1\,\Phi_2,\, A^2,\, B^2,\, A\,B,\, A\,\Phi_1,\, A\, \Phi_2,\, B\,\Phi_1,\, B\, \Phi_2\}\, \cdots\}
  \end{equation}
while at $k=2$ we would have

\begin{equation}
\{ \{1\},\, \{\Phi_1,\,\Phi_2\},\,\{\Phi_1^2,\, \Phi_2^2,\, \Phi_1\,\Phi_2,\, A^2,\, B^2,\, A\,B\}\, \cdots\}
  \end{equation}
Thus, we can explicitly see how the number of operators matches each power of $t$ in each case. Note that, for finite $k$, at the non-abelian level, the operators counted would involve monopole operators. 

In order to avoid relying on the properties of the monopole operators, we are interested in the large $k$ limit. Since $t$ is a complex number such that $|t|<1$, in the large $k$ limit $t^k\rightarrow 0$, and thus the Hilbert series coincides with that of weighted $\mathbb{C}^3$

\begin{equation}
g_{k\rightarrow \infty\sim \mathbb{C}^3_{2,1,1}}=\frac{1}{(1-t^2)\, (1-t)^2}
\end{equation}
The $\mathbb{C}^3$ has the same weight for two out of the three directions, while the third is weighted twice as much. This is actually compatible with the R-charge assignation we suggested above. In the large $k$ limit, the effective $\mathbb{C}^3$ is spanned by $z_1=\Phi_1,\, z_2=\Phi_2\,, m=A\,B$. As discussed, the global $SU(2)$ symmetry ensures that $\Phi_i$ have the same R-charge $R_{\Phi}$. Thus, in order for the superpotential to have R-charge 2, $R_{m}=2\, (1-R_{\Phi})$. The Hilbert series above requires that $R_{m}=2\, R_{\Phi}$, which leads to $R_{\Phi}=1/2$, as anticipated above.

\section{On the $SO(4)$ symmetry}

So far, we have seen that the mesonic moduli space of the HVZ theory contains the orbifold $\mathbb{C}^2\times \mathbb{C}^2/\mathbb{Z}_k$. Also, the coherent branch of the chiral ring matches the functions on this geometry. However, this orbifold preserves $\mathcal{N}=4$ SUSY, and as such we expect $SO(4)$ R-symmetry. Following \cite{Aharony:2008ug} and \cite{Benna:2008zy} we might try to search for such a symmetry in the bosonic potential of the theory. However, it is easy to realize that this is not the case. After integrating out the $D_g$, the bosonic potential (\ref{V}) reduces to

\begin{equation}
\label{VV}
V_b={\rm Tr}\Big(- \sum_{X_{ab}}(\sigma_aX_{ab}-X_{ab}\sigma_b)(\sigma_aX_{ab}-X_{ab}\sigma_b)^{\dagger}-\sum_{X_{ab}}|\partial_{X_{ab}}W|^2\Big)
\end{equation}
where

\begin{equation}
\mu_g=4\,k_g\, \sigma_g
\end{equation}
with the moment maps given by (\ref{momap}), while the F-terms are (\ref{F}). After substituting all the ingredients, the bosonic potential will contain terms quadratic and quartic in the bifundamentals, as well as terms with no bifundamentals and only adjoints. Let us look at those, which can only come from the first term in (\ref{VV}). Indeed, all the terms sextic in the adjoints are

\begin{equation}
\label{Vadjoint}
V_b\supset -\frac{1}{16\, k^2}\, {\rm Tr}\Big(\, [\Phi_j,[\Phi_i,\Phi_i^{\dagger}]]\,[[\Phi_i,\Phi_i^{\dagger}],\Phi_j^{\dagger}]\Big)
\end{equation}
where repeated indices sum. 

From our previous findings, we know that the $SU(2)$ symmetry which acts on the $\Phi_i$ fields is part of the global $SO(4)$ symmetry we are after. Indeed, we would expect that upon writing

\begin{equation}
\label{cartesian}
\Phi_1=X_1+i\, X_2\ , \qquad \Phi_2=X_3+i\, X_4
\end{equation}
the $SO(4)$ would appear as a rotation of the $X_I$ fields. However, after a bit of algebra, one can convince oneself that such $SO(4)$ is not present in (\ref{Vadjoint}). \footnote{The investigation of the bosonic potential of the HVZ theory leading to these conclusions was initiated in collaboration with M.Benna and F.Benini.} It is natural to conjecture that the non-appearance of the $SO(4)$ is connected with the seemingly extra branch in the chiral ring at the origin of the orbifold; which in turn seems also naturally connected with the index mismatch.  

We can also ask about the fate of the expected $SU(2)$ isometry of the orbifolded $\mathbb{C}^2$ which should appear as a global symmetry of the dual theory. After some algebra, one can convince oneself that such a symmetry is also not present in the full bosonic potential, since the structure of terms with adjoints and bifundamentals fails to furnish an $SU(2)$ invariant. However, as it will turn out to be useful later on, the sextic terms in the bifundamentals alone do indeed admit such an $SU(2)$ action.

In order to perform yet another test, we now explore a class of solutions of the HVZ theory which should exhibit the $SO(4)$ invariance. Recall that the mesonic moduli space is $\mathbb{C}^2\times \mathbb{C}^2/\mathbb{Z}_k$. If we go to cartesian coordinates, inside the first $\mathbb{C}^2$ factor there is an $S^3$, while inside the second factor there is an orbifolded $S^3$. Thus, we expect that the spike solutions of this theory fall into two classes: those expanding in the $S^3$ and those expanding in the orbifolded $S^3$. Interestingly, only the latter class would feel the $\mathbb{Z}_k$ orbifold. It is then natural to expect that this class is analogous to the spike found in \cite{Terashima:2008sy, Gomis:2008vc}, which has been shown to be a non-commutative version of the Hopf fibering \cite{Nastase:2009ny}. In turn, the first class should correspond to fuzzy $S^3$-like spikes along the lines of \cite{Guralnik:2000pb,Ramgoolam:2001zx,Ramgoolam:2002wb}, which should reduce, in the large $k$ limit, to a $NS5-D2$ intersection, as opposed to a $D2-D4$ intersection. It is precisely this $S^3$ what is rotated by the $SO(4)$ in which we are interested on.

\subsection{BPS solutions in $\mathbb{C}^2$}

Since we are interested on M5 growing in the $\mathbb{C}^2$ not acted by the $k$-orbifold, it is natural to consider $A=B=0$. It is then clear that the bosonic potential reduces to just (\ref{Vadjoint}). We can now consider configurations where $\Phi_i=\Phi(\sigma)$, being $\sigma=x_2$, while all the gauge fields are vanishing. After a straightforward manipulation, the action becomes

\begin{equation}
S=-\int\, |\partial_{\sigma}\Phi_j\pm\frac{1}{4k}[[\Phi_i,\Phi_i^{\dagger}],\Phi_j]|^2\mp\frac{1}{8k}\int \partial_{\sigma}([\Phi_i,\Phi_i^{\dagger}])^2
\end{equation}
Thus, as long as

\begin{equation}
\partial_{\sigma}\Phi_j\pm\frac{1}{4\,k}[[\Phi_i,\Phi_i^{\dagger}],\Phi_j]=0
\end{equation}
the action becomes a total derivative and therefore we have a stable BPS solution. 

Let us discuss the consistency of our truncation of the action. Since the bifundamental fields should appear always quadratically in the action, it is clear that the ansatz $A=B=0$ satisfies their equations of motion. Furthermore, we have to make sure that Gauss law for the gauge fields is satisfied. For the gauge field in node 1, since $A,\, B$ are vanishing, it is clear that its equations motion are trivially satisfied assuming that it vanishes. In turn, the equations of motion for the gauge-field in node 2 read

\begin{equation}
\frac{k_2}{2\pi}\, \epsilon_{\mu\nu\rho}\, F_{(2)}^{\mu\nu\rho} = \Phi_i^{\dagger}\, D_{\mu}\Phi_i-\Phi_i\, D_{\mu}\Phi_i^{\dagger}+D_{\mu}\Phi_i^{\dagger}\,\Phi_i-D_{\mu}\Phi_i\, \Phi_i^{\dagger}
\end{equation}

Thus, setting $A_{(2)}=0$ we can satisfy the above condition by taking $\partial_{\sigma}\Phi_i=f^{-1}\, f'\, \Phi_i$ with $f$ a real function, that is, we consider

\begin{equation}
\Phi_1=f(\sigma)\, (G_1+iG_2)\ ,\qquad \Phi_2=f(\sigma)\, (G_3+iG_4)
\end{equation}
being $G_i$ four hermitian constant matrices and $f\in \mathbb{R}$. In this language, the cartesian coordinates (\ref{cartesian}) are $X_i=f(\sigma)\, G_i$. 

Since from the geometrical point of view we expect an spike with $S^3$ profile, we might have hoped that the $G_i$ are the matrices corresponding to a fuzzy $S^3$ as described in \cite{Guralnik:2000pb,Ramgoolam:2001zx,Ramgoolam:2002wb}. However, it turns out that these matrices do not solve (except for $n=1$, that is, $G_i$ being the $SO(4)$ Dirac matrices; see appendix) the BPS equations. In turn, if we consider the $G_i$ to be identified with the first 4 matrices of the fuzzy $S^5$ as described in  \cite{Castelino:1997rv} we have

\begin{equation}
[[\Phi_i,\Phi_i^{\dagger}],\Phi_j]=8\, f^2\, \Phi_j
\end{equation}
so that we are left with 

\begin{equation}
\partial_{\sigma}f\pm \frac{2}{k}\, f^3=0\qquad \rightarrow \qquad f^2=\pm \frac{k}{4\sigma}
\end{equation}

Furthermore, using the properties of the $G$ matrices (see appendix), we can see that

\begin{equation}
{\rm Tr}\Big([\Phi_i,\,\Phi_i^{\dagger}]\Big)^2=8 f^4\, {\rm Tr}\,\sum_{i=1}^4G_iG_i
\end{equation}

It is now natural to ask to what extend is this a fuzzy $S^3$ as expected. Being a subset of the matrices of a fuzzy $S^5$, the $G_i$ can be seen as endomorphisms of a space $\mathbb{V}$ whose dimension is written in terms of an integer $n$ as $N=\frac{(n+1)(n+2)(n+3)}{6}$ \cite{Castelino:1997rv,Guralnik:2000pb,Ramgoolam:2001zx,Ramgoolam:2002wb}. As reviewed in the appendix, this space decomposes as the direct sum of $SO(4)$ representations $V_{(q,\, n-q)}$ as

\begin{equation}
\mathbb{V}=\oplus_{q=0}^n V_{(q,\,n-q)}
\end{equation}

Using the properties of the $G_i$ matrices (see appendix), it turns out that $\sum_1^4\, G_i^2$ is a block diagonal matrix whose non-vanishing blocks are proportional to the identity, but with a different proportionality constant $r_i$ for each irrep. of $SO(4)$ (that is, for each $V_{(q,\,n-q)}$). Then, we can write

\begin{equation}
\sum_1^4\, X_i^2=\left(\begin{array}{c c c} r_1\, \unity_{dim(r_1)} & 0 & 0 \\ 0 & r_2\,  \unity_{dim(r_2)} & 0 \\ 0 & 0 & \ddots \end{array}\right)
\end{equation}
 Thus, it is natural to interpret this as a collection of stacks of  $M5$ branes, each with radius $r_i$.  Then, the central charge, which reads

\begin{equation}
Z=\frac{1} {k}\, {\rm Tr}\,\sum_{i=1}^4G_iG_i\,\int d^2x_{1,2}\, df^4\ ,
\end{equation}
becomes in terms of the $r_i$'s 

\begin{equation}
Z=\sum_{q=0}^{n}\,\frac{{\rm Tr}\,\unity_{(q)}}{k\, \rho_q^2}\int d^2x_{1,1}\, d r_q^4\ .
\end{equation}

Introducing the angular integration on an $S^3$, and changing to cartesian coordinates, we can naturally re-write the central charge as

\begin{equation}
Z=\sum_{q=0}^{n}\,\frac{16}{\pi\,k}\,\frac{{\rm Tr}\,\unity_{(q)}}{\rho_q^2}\int d^6x_q\ ,
\end{equation}
where we have taken into account the extra $1/8\pi$ hidden in our definition of $k$ with respect to the conventions in \cite{Aharony:2008ug}. Recalling now that in units where $T_2=1$ we have $T_5=1/2\pi$

\begin{equation}
Z=\sum_{q=0}^{n}\,\frac{32}{k}\,\frac{{\rm Tr}\,\unity_{(q)}}{\rho_q^2}\,T_5\,\int d^6x_q
\end{equation}
At $k=1$ this  should be a collection of stacks of $M5$ branes. It is then natural to define the number of branes in each stack as

\begin{equation}
n_q=32\,\frac{{\rm Tr}\,\unity_{(q)}}{\rho_q^2}
\end{equation}
such that

\begin{equation}
Z=\sum_{q=0}^{n}n_q\,T_5\,\int d^6x_q
\end{equation}
becomes the energy for a collection of stacks of $n_q$ M5 branes. Also, using the formulae in the appendix

\begin{equation}
n_q=16+\frac{16\,(1-n)}{2n+q\,n-q^2}
\end{equation}

The interpretation of the spike as a brane makes only sense in the large $n$ limit. In that case, we can approximate the above expression by

\begin{equation}
n_q=16-\frac{16}{2+q-\frac{q^2}{n}}
\end{equation}
As a rough estimate, one can compute the average value of $n_q$ as $n$ grows and verify that $\langle n_q\rangle \rightarrow 16$. Thus, since $n_q\le 16\,\,\forall q$, in the large $n$ limit, the vast majority of the representations have $n_q=16$. However, the significance of $n_q$ is an issue which remains to be clarified. 

For large $k$, in turn, a IIA description becomes more suitable. It is convenient to recall that the central charge comes from a three-dimensional action which we re-wrote, upon evaluation in the action, as

\begin{equation}
Z=\sum_{q=0}^{n}\,n_q\,k^{-1}\,T_5\,\int d^2x_{1,1}\, dr_q\, r_q^3
\end{equation}
In particular, we stress that, after changing the integration variable to $df^4$, this still carries a hidden factor of the 2-dimensional M-theoretic induced metric, which in this case is just one. However, when going to IIA, the string frame metric reads in terms of the 11d metric as $G_{\mu\nu}^{11d}=e^{-\frac{2\Phi}{3}}g_{\mu\nu}^{IIA_{string}}$. Thus, this re-scaling introduces a factor of $e^{\frac{2\Phi}{3}}$. In turn, we expect that $e^{\frac{2\Phi}{3}}=k^{-1}$, so in IIA\footnote{One way to see that $e^{\frac{2\Phi}{3}}=k^{-1}$ is recalling that the orbifolded angle appears in the 11d metric as $ds^2= d\psi^2+\cdots$. Since $\psi\in [0,2\pi/k]$, we can do $k^{-2}\,d\tilde{\psi}^2$, with $\tilde{\psi}=k\,\psi\in[0,2\pi]$ and use the standard reduction formulas along $\tilde{\psi}$. It is then clear that the radius of the circle is $k^{-1}$, which leads to $e^{\frac{2\Phi}{3}}=k^{-1}$.}

\begin{equation}
Z_{IIA}=\sum_{q=0}^{n}\,n_q\, e^{2\Phi}\, T_5\,\int d^6x_q^{IIA}
\end{equation}
which, since $e^{2\Phi}\, T_5=T_{NS5}$ matches the expected picture of this spike representing a D2/NS5 intersection.

\subsection{BPS solutions in $\mathbb{C}^2/\mathbb{Z}_k$}

We could also consider solutions where $\Phi_i$  and both gauge fields vanish while the $A, B$ fields are non-vanishing. As in the previous subsection, it is straightforward to check that this is a consistent truncation of the action. We expect these solutions to correspond to $M5$ branes growing in the $\mathbb{C}^2/\mathbb{Z}_k$ part of the geometry. In this case, it is easy to see that

\begin{equation}
V_b=-\frac{1}{16\, k^2}{\rm Tr}\Big(\, |A\, B\, B^{\dagger}-B^{\dagger}\, B\, A|^2+|B\, A\, A^{\dagger}-A^{\dagger}\, A\, B|^2\Big)
\end{equation}
Thus, the action can be written as

\begin{eqnarray}
S=&&-\int \, |\frac{dA}{d\sigma}-\frac{1}{4\,k}(A\, B\, B^{\dagger}-B^{\dagger}\, B\, A)|^2+ |\frac{dB}{d\sigma}-\frac{1}{4\,k}(A^{\dagger}\, A\, B-B\, A\, A^{\dagger})|^2\nonumber \\ &&+\frac{1}{4\,k}\,\int \frac{d}{d\sigma}\Big(A\, A^{\dagger}\,B^{\dagger}\, B-B\, B^{\dagger}\, A^{\dagger}\, A\Big)
\end{eqnarray}
where for simplicity we choose one sign when completing squares. Thus, we have the BPS equations

\begin{equation}
\frac{dA}{d\sigma}-\frac{1}{4\,k}(A\, B\, B^{\dagger}-B^{\dagger}\, B\, A)=0\ ,\quad \frac{dB}{d\sigma}-\frac{1}{4\,k}(A^{\dagger}\, A\, B-B\, A\, A^{\dagger})=0
\end{equation}
We can define now $C_i=(A,B^{\dagger})$. Then, the equations can be compactly written as

\begin{equation}
\frac{dC_i}{d\sigma}-\frac{1}{4\, k}(C_i\, C_j^{\dagger}\, C_j-C_j\, C_j^{\dagger}\, C_i)=0
\end{equation}
whose solutions are well-known\cite{Terashima:2008sy, Gomis:2008vc} and yield to an M2-M5 intersection whose IIA interpretation is a D2-D4 intersection \cite{Hanaki:2008cu,Nastase:2009ny}. Note in particular the appearance of the $SU(2)$ invariance in the $BPS$ equations of motion corresponding to the $SU(2)$ isometry of the $\mathbb{C}^2/\mathbb{Z}_k$ orbifold. We would like to stress, however, that as for the $SO(4)$, this symmetry is not present in the full bosonic potential, $\textit{i.e.}$ once the mixing of F and D terms is taken into account. 

Indeed, this spike \textit{$\grave{a}$ la}  \cite{Terashima:2008sy, Gomis:2008vc} can be understood in a rather generic fashion. We first note that it comes from mixing derivative terms with the D-term potential. Thus, as long as we can consistently set to zero all F-terms by turning on only two fields (which is automatic if $W$ is of degree 4 or higher in the fields; or if no adjoint is present in the two adjacent gauge nodes involved); as long as we have a 3d Chern-Simons/matter theory whose quiver involves two arrows between two adjacent nodes, it  is always possible to find a spike solution along the lines of the one discussed here. A similar statement is also true for the fuzzy $S^3$ ones when two adjoints are present on a node, as some of the theories in \cite{Benishti:2009ky} satisfy. 

\section{Matching operators with SUGRA fluctuations}

Even though the full $SO(4)$ is not present in the bosonic potential of the theory, we have succeeded in matching chiral operators with the expected spectrum of functions and found first hints of the appearance of the $SO(4)$ at least in the BPS solutions. With these hints, we will assume that at least on the coherent branch, the desired $SO(4)$ appears. It is then natural to work with the $X_I$ fields in (\ref{cartesian}) instead of the $\Phi_i$ fields. Then, among all possible operators we might consider, we can restrict ourselves to those with no index on the $\mathbb{C}^2/\mathbb{Z}_k$ factor; that is, operators constructed as products of only $X_I$'s.

\subsection{SUGRA excitations}

We are interested in the spectrum of 11d SUGRA fluctuations in the near-horizon limit of M2 branes in $\mathbb{C}^2\times \mathbb{C}^2/\mathbb{Z}_k$. At this point we can actually be generic and consider M2 branes moving in a $CY_4$ cone constructed starting from a smaller conical $CY_{4-n}$ by adding the suitable factors of $\mathbb{C}$, $i.e.$ $\mathbb{C}^n\times CY_{4-n}$. We should note that particular results for some $CY_{4-n}$ of the formulas presented in this subsection have been used recently in similar contexts in $e.g.$ \cite{Ahn:2009bq}.

Given the $CY_{4-n}$ property, there is a holomorphic top form $\Omega_{4-n}$ on $CY_{4-n}$ globally well-defined. If we parametrize $\mathbb{C}^n$ with a set of $n$ complex coordinates $z_i$, we can construct the following holomorphic top form in the product space

\begin{equation}
\Omega_4=dz_1\wedge \cdots \wedge dz_n\wedge \Omega_{4-n}
\end{equation}
which is clearly well-defined everywhere, so that the product space is also Calabi-Yau.

We will assume the $CY_{4-n}$ to be a (generically singular) cone over a real dimension $7-2n$ base $b$. In turn, the $\mathbb{C}^n$ factor is the same as $\mathbb{R}^{2n}$, and thus we can write it as a non-singular cone over $S^{2n-1}$. Thus, we can write the metric of the $CY_4$ as

\begin{equation}
ds^2=dr_1^2+r_1^2 \,ds_b^2+dr_2^2+r_2^2 \,d\Omega_{2n-1}^2
\end{equation}
By defining

\begin{equation}
r_1=r\cos\beta\ ,\qquad r_2=r\sin\beta
\end{equation}
the metric becomes of the standard cone-like form

\begin{equation}
ds^2=dr^2+r^2\, ds_B^2\ ,\qquad ds_B^2=d\beta^2+\sin^2\beta\, d\Omega_{2n-1}^2+\cos^2\beta\, ds_b^2
\end{equation}
Given the cone structure and the $CY_4$ property, we can place a stack of $N$ M2 branes in this space thus leading to an, at least, $\mathcal{N}=2$ SCFT dual to the near horizon geometry\footnote{Recall that, in order to obtain the $AdS_4\times B$ background we need to re-define the radial coordinate $r^2=2\rho$.}

\begin{equation}
ds^2=\frac{\rho^2}{\mathcal{R}^2/4}dx_{1,2}^2+\frac{\mathcal{R}^2/4}{\rho^2}d\rho^2+\mathcal{R}^2ds_B^2
\end{equation}
Among all possible 11d SUGRA fluctuations, we will consider a minimally coupled scalar which can be thought as a metric fluctuation corresponding to a graviton polarized along the $AdS_4$ (see $e.g.$ section 4 of \cite{Klebanov:2009kp}). The free field equation of motion reduces to

\begin{equation}
\Box_{AdS_4}\,\phi+\frac{1}{4\mathcal{R}^2}\, \mathbf{A}\, \phi=0\ ;
\end{equation}
where $\mathbf{A}$ is the angular laplacian on $B$. Let $\frac{1}{\mathcal{R}^2}\, \mathbf{A}=-m^2\, \phi$ so that $m$ is the mass as seen from $AdS_4$.  Thus, we see that we need the eigenvalues of the laplacian in $CY_4$

\begin{equation}
\mathbf{A}\, \phi=-E_I\, \phi\, 
\end{equation}
since $m^2\mathcal{R}^2=E_I$. Then, the conformal dimension of the dual field is given by

\begin{equation}
\Delta=\frac{3}{2}+\frac{1}{2}\sqrt{9+m^2\mathcal{R}^2}\ .
\end{equation}

After some algebra, the laplacian in $B$ reads

\begin{equation}
\mathbf{A}\phi=\frac{1}{ \sin^{2n-1}\beta\cos^{7-2n}\beta}\, \partial_{\beta}\Big(\sin\beta\cos^5\beta\,\partial_{\beta} \phi\Big)+\frac{1}{ \sin^2\beta}\mathbf{A}_{S^{2n-1}} \phi+\frac{1}{\cos^2\beta}\, \mathbf{A}_b \phi\ ;
\end{equation}
being $\mathbf{A}_b$ the laplacian on $b$ and $\mathbf{A}_{S^{2n-1}}$ the laplacian on the $S^{2n-1}$. 

Let us suppose the eigenproblem in $b$ to be known, $i.e.$

\begin{equation}
\mathbf{A}_bf_I=-\mathcal{E}_If_I\ .
\end{equation}
Then, we can construct the eigenfunctions as $\phi=Y_{S^{2n-1}}^l\, f_I(b)\,\psi(\beta)$, being $Y_{S^{2n-1}}^l$ a spherical harmonic of $S^{2n-1}$ with eigenvalue $-l(l+2n-2)$ provided that $\psi$ satisfies

\begin{equation}
\frac{1}{ \sin^{2n-1}\beta\cos^{7-2n}\beta}\, \partial_{\beta}\Big(\sin^{2n-1}\beta\cos^{7-2n}\beta\,\partial_{\beta} \psi\Big)-\Big(\frac{l(l+2n-2)}{ \sin^2\beta}+\frac{\mathcal{E}_I}{\cos^2\beta}-E_I\Big)\psi=0\ .
\end{equation}
It is convenient to introduce $u=\cos^2\beta$. Then, the above equation reduces to

\begin{equation}
\label{geneq}
u\, (1-u)\, \partial_u^2\psi+(4-n-4u)\, \partial_u\psi-\Big(\frac{\frac{l(l+2n-2)}{4}}{ 1-u}+\frac{\frac{\mathcal{E}_I}{4}}{u}-\frac{E_I}{4}\Big)\psi=0\ .
\end{equation}
\subsection{The special case of $l=0$}

For generic $\mathcal{E}_I$, $l$ and $n$ (\ref{geneq}) is too complicated. Let us focus on the sector with $l=0$, that is, on fluctuations which only see the $CY_{4-n}$. In that case, the equation becomes much simpler

\begin{equation}
u\, (1-u)\, \partial_u^2\psi+(4-n-4u)\, \partial_u\psi-\Big(\frac{\frac{\mathcal{E}_I}{4}}{u}-\frac{E_I}{4}\Big)\psi=0
\end{equation}
The solutions come in terms of hypergeometric functions as

\begin{equation}
\psi_1=u^{\frac{1}{2}\big(n-3+\sqrt{\mathcal{E}_I+(n-3)^2}\big)}\, _2F_1(a_1,b_1,c_1;u)\ ;
\end{equation}
with
\begin{equation}
a_1=\frac{1}{2}\Big(-\sqrt{9+E_I}+\sqrt{\mathcal{E}_I+(n-3)^2}+n\Big)\ ,\quad 
b_1=\frac{1}{2}\Big(\sqrt{9+E_I}+\sqrt{\mathcal{E}_I+(n-3)^2}+n\Big)
\end{equation}
\begin{equation}
c_1=1+\sqrt{\mathcal{E}_I+(n-3)^2}\ ;
\end{equation}
and 

\begin{equation}
\psi_2=u^{\frac{1}{2}\big(n-3-\sqrt{\mathcal{E}_I+(n-3)^2}\big)}\, _2F_1(a_2,b_2,c_2;u)
\end{equation}
with
\begin{equation}
a_2=\frac{1}{2}\Big(-\sqrt{9+E_I}-\sqrt{\mathcal{E}_I+(n-3)^2}+n\Big)\ ,\, b_2=\frac{1}{2}\Big(\sqrt{9+E_I}-\sqrt{\mathcal{E}_I+(n-3)^2}+n\Big)\end{equation}
\begin{equation}
c_2=1-\sqrt{\mathcal{E}_I+(n-3)^2}\ .
\end{equation}

One can check that the regular solution is $\psi_1$. Furthermore, in order to have a finite order polynomial, we need to impose that $a_1=-j$ with $j\in \mathbb{N}$, which leads to

\begin{equation}
E_I=\Big(2j+n+\sqrt{(n-3)^2+\mathcal{E}_I}\Big)^2-9\ .
\end{equation}
Therefore, the dimension of the corresponding dual operators is

\begin{equation}
\label{diml=0}
\Delta=\frac{3}{2}+j+\frac{n}{2}+\frac{1}{2}\sqrt{(n-3)^2+\mathcal{E}_I}\ .
\end{equation}

\subsection{The special case with $\mathcal{E}_I=0$ and a consistency check}

We can alternatively consider fluctuations which do not see the $CY_{4-n}$ by  concentrating on the sector with $\mathcal{E}_I=0$. The equation becomes now

\begin{equation}
u\, (1-u)\, \partial_u^2\psi+(4-n-4u)\, \partial_u\psi-\Big(\frac{\frac{l(l+2n-2)}{4}}{ 1-u}-\frac{E_I}{4}\Big)\psi=0\ .
\end{equation}

In this case the solutions are

\begin{equation}
\psi_1=u^{n-3-\sqrt{(n-3)^2+l(l+2n-2)}}\, _2F_1(a_1,b_1,c_1; u)\ ,
\end{equation}
where 
\begin{equation}
a_1=\frac{1}{2}\Big(n-\sqrt{9+E_I}-\sqrt{(n-3)^2+l(l+2n-2)}\Big)\, ,
\end{equation}

\begin{equation}
b_1=\frac{1}{2}\Big(n+\sqrt{9+E_I}-\sqrt{(n-3)^2+l(l+2n-2)}\Big)\ ,
\end{equation}
\begin{equation}
c_1=1-\sqrt{(n-3)^2+l(l+2n-2)}\ ;
\end{equation}
and
\begin{equation}
\psi_2=u^{n-3+\sqrt{(n-3)^2+l(l+2n-2)}}\, _2F_1(a_2,b_2,c_2; u)\ ,
\end{equation}
where now
\begin{equation}
a_2=\frac{1}{2}\Big(n-\sqrt{9+E_I}+\sqrt{(n-3)^2+l(l+2n-2)}\Big)\ ,
\end{equation}

\begin{equation}
 b_2=\frac{1}{2}\Big(n+\sqrt{9+E_I}+\sqrt{(n-3)^2+l(l+2n-2)}\Big)\ ,
\end{equation}
\begin{equation}
c_1=1+\sqrt{(n-3)^2+l(l+2n-2)}\ .
\end{equation}

The regular solution is now $\psi_2$. Imposing again $a_2=-j$ we have

\begin{equation}
E_I=\Big(2j+n+\sqrt{9-2l+l^2-6n+2ln+n^2)^2}\Big)^2-9\ .
\end{equation}
The dimension of the corresponding dual operators is

\begin{equation}
\label{dimE=0}
\Delta=\frac{3}{2}+j+\frac{n}{2}+\frac{1}{2}\sqrt{9-2l+l^2-6n+2ln+n^2)}\ .
\end{equation}

\vspace{1cm}

We can now focus on the case of interest, namely the $\mathbb{C}^2\times \mathbb{C}^2/\mathbb{Z}_k$ space. Let us start by considering the case $k=1$, where there is an obvious $\mathbb{Z}_2$ symmetry exchanging the two $\mathbb{C}^2$. Since in that case both factors are cones over $S^3$, this discrete symmetry just exchanges the two base $S^3$. Thus, we should expect that in this case both (\ref{diml=0}) and  (\ref{dimE=0})  coincide. Indeed, one can see that taking $n=2$ in (\ref{dimE=0}) precisely coincides with (\ref{diml=0}) upon replacing $\mathcal{E}_I=l\,(l+2)$.

Coming back to the case of interest, we will consider that in the $\mathbb{C}^2/\mathbb{Z}_k$ orbifold is playing the role of the $CY$. Thus, since we are interested in operators in the large $k$ limit which do not see the orbifolded $\mathbb{C}^2$, we should take $\mathcal{E}_I=0$ above. Thus, for those operators, the conformal dimensions are

\begin{equation}
\label{dim}
\Delta=3+j+\frac{l}{2}
\end{equation}

\subsection{Field theory operators}

As stated, we will concentrate on the subsector of operators of the coherent branch which only see the $\mathbb{C}^2$ not acted by the higher $k$ orbifold. Thus, we are led to consider operators constructed only from $\Phi_i$ fields in the coherent branch. As discussed, in terms of these we only see an explicit $SU(2)$. In order to see the full $SO(4)$ we need to work with the $X_I$. Motivated by the one to one correspondence with functions in the variety plus the BPS intersections, we will assume that, in the coherent branch, the operators of interest can be written as

\begin{equation}
\mathcal{O}_{I_1\cdots I_l}={\rm Tr}\, \prod X_{\{I_1}\cdots X_{i_l\}_{\rm Traceless}}\, ;
\end{equation}
where the notation $\{\cdots\}_{{\rm Traceless}}$ stands for symmetrized and traceless $SO(4)$ indices, so that the operator has integer spin $l$. Following our assumption, now also supported by the Hilbert series matching, that the $R_{\Phi}\frac{1}{2}$, we have that the dimension of these operators is then $\frac{l}{2}$. 

We can now use these harmonics to construct the operators

\begin{equation}
\label{O}
\mathcal{O}_{\mu\nu;\, I_1\cdots I_l}={\rm Tr}\, \mathcal{T}_{\mu\nu}\, \mathcal{O}_{I_1\cdots I_l}
\end{equation}
being $\mathcal{T}_{\mu\nu}$ the stress-energy tensor. These operators are in a $l$ representation of $SO(4)$ and have dimension $3+\frac{l}{2}$. 

More generically, we could also consider operators of the form

\begin{equation}
\label{Oj}
\mathcal{O}_{\mu\nu;\, I_1\cdots I_l|\, j}={\rm Tr}\, \mathcal{T}_{\mu\nu}\, \mathcal{O}_{I_1\cdots I_l}\, P_j(\vec{X}^2)
\end{equation}
being $P_j$, at this point, an operator (non-chiral primary) of dimension $j$. Such operator should be an order $j$ polynomial in $\vec{X}^2$.  

We would like now to find the dual gravitational modes. Since we are considering operators with insertions of the stress-energy tensor, we expect them to be dual to spin 2 metric fluctuations polarized along the $AdS_4$, that is, $AdS_4$ gravitons. Therefore, the dual fields should be among the modes investigated in the previous subsection. As noted in the previous subsection, at higher $k$ the modes not seeing the orbifolded part of the space will have $\mathcal{E}_I=0$, which leads us to 

\begin{equation}
\Delta=3+j+\frac{l}{2}
\end{equation}

For $j=0$ we obtain a matching with the dimension and quantum numbers of the operators (\ref{O}) .

For $j\ne 0$ we will have contributions from the insertion of the $P_j$ polynomial of dimension $j$ in $\vec{X}^2$. This operator is dimension $j$ and thus gives a dimension $3+j+\frac{l}{2}$. We propose those operators to be the ones of the form (\ref{Oj}). 

The wavefunctions for generic $j$ become

\begin{equation}
\psi=u^{\frac{l}{2}}\, _2F_1(-j,3+j+l,2+l;u)
\end{equation}
From here we can actually fix the polynomial $P_j$ above, since the hypergeometric function becomes just a Legendre polynomial $P_j$. To illustrate, we can read a few operators and their quantum numbers

\begin{center}
\begin{tabular}{|c| c| c| c|}
\hline
Operator & $SO(4)$ & $j$ & $\Delta$ \\
\hline
$ {\rm Tr}\, \mathcal{T}_{\mu\nu}\, X_I$ & $\mathbf{4}$ & 0 & $ \frac{5}{2}$\\
\hline
$ {\rm Tr}\, \mathcal{T}_{\mu\nu}\, X_I\,X_J$ & $\mathbf{6}$ & 0 & $ 4$\\
\hline \hline
$ {\rm Tr}\, \mathcal{T}_{\mu\nu}\, X_I\, (1-\frac{5}{3}\, \vec{X}^2)$ & $\mathbf{4}$ & 1 & $ \frac{7}{2}$\\
\hline
$ {\rm Tr}\, \mathcal{T}_{\mu\nu}\, X_I\,X_J\, (1-\frac{3}{2} \vec{X}^2)$ & $\mathbf{6}$ & 1 & $ 5$\\
\hline
\end{tabular}
\end{center}

\section{Conclusions}

In this note we have concentrated on various aspects of the HVZ theory. This theory has as mesonic moduli space $\mathbb{C}^2\times \mathbb{C}^2/\mathbb{Z}_k$. This orbifold preserves $\mathcal{N}=4$ SUSY, which suggests that the theory should involve such SUSY. If so, an $SO(4)_R$ symmetry is expected. Moreover, an extra $SU(2)$ global symmetry remanent of the broken $SO(4)$ by the orbifold is expected. However, neither this global symmetry, nor the $SO(4)_R$ one are present in the bosonic potential of the theory. 

We investigated the chiral operators of the theory. For the ones living in the coherent branch of the moduli space, we were able to find perfect matching with the KK harmonics from the geometry point of view. Nevertheless, at the origin of the orbifolded $\mathbb{C}^2$, an extra branch of dimension $\mathcal{O}(N^2)$ for the chiral ring appears. At the same time,  a small mismatch in the superconformal index between SUGRA and field theory has been found in \cite{Choi:2008za}. It seems natural to expect that these two facts are connected. It would be interesting to investigate this further.

Even though possible present in the computation of the abelian moduli space, the extra branches do not show up in the chiral ring in other examples, such as $Q^{1,1,1}$ or $M^{3,2}$. The meaning of such extra branches remains an important question to clarify for an interpretation of the HVZ theory as describing M2 branes.

Focusing on the coherent branch, inspired by the explicit matching of chiral operators and KK harmonics at large $k$ and the BPS solutions with the expectations for a $\mathbb{C}^2\times \mathbb{C}^2/\mathbb{Z}_k$ moduli space, we conjectured that on this branch it is actually possible to write the chiral operators in a manifestly $SO(4)$ invariant way. By focusing on certain operators involving insertions of the stress energy tensor together with chiral primaries which do not see the orbifolded part of the moduli space, we found a satisfactory match with the spin 2 metric fluctuations polarized along the $AdS_4$. In particular, the R-charge assignation is consistent with that expected from inspection of the moduli space.

Interestingly, we succeeded in constructing \textit{transverse} M5 branes, \textit{i.e.} fivebranes which do not wrap the M-theory circle. We believe this is the first construction in the literature coming from a motivated lagrangian. However, our construction raises many questions. First of all, it is not exactly a fuzzy $S^3$ as in \cite{Guralnik:2000pb,Ramgoolam:2001zx,Ramgoolam:2002wb}. Instead, it looks more like a collection of different radii $S^3$. It would be nice to clarify the properties of such construction. In particular, it would be interesting to compute the spectrum of linearized fluctuations to see wether it matches that of a 3-sphere. Also, the quantity $n_q$ should be better understood. 

It would be interesting to study possible superconformal mass-deformations of the HVZ theory along the lines of \cite{Gomis:2008vc}. In that case, one would expect that the discrete set of vacua of such theory are very similar to the spike solutions which we have found. Since the size of the matrices grows very differently for the spikes \textit{$\grave{a}$ la} \cite{Terashima:2008sy, Gomis:2008vc} as opposed to those \textit{$\grave{a}$ la} Castelino, Lee \& Taylor, the counting in \cite{Gomis:2008vc} might me affected (indeed, naively it looks it would help matching field theory with SUGRA, since the fuzzy $S^3$ spikes are much less than the other ones in the large $N$ limit).

Very recently, an inverse algorithm to construct dual theories for M2 branes probing arbitrary $CY_4$ has been started being devised in \cite{Davey:2009bp,Hewlett:2009bx,Davey:2009qx, Benishti:2009ky,Taki:2009wf}. In particular, in \cite{Benishti:2009ky}, a plethora of theories with moduli space $\mathbb{C}^2\times \mathbb{C}^2/\mathbb{Z}_2$ and $\mathbb{C}^4$ at level $k=1$ has been found. These theories are a bit mysterious in that they involve vanishing CS levels and non-standard multiplicities for the perfect matchings in the toric diagram.\footnote{Also, as already pointed out in \cite{Benishti:2009ky}, as suggested in \cite{Imamura:2009ph}, for non-chiral theories dual to M2 probing $\mathcal{C}(B)$, the number of nodes should be $2+{\rm dim}\,H_2(B)$, that is, 2 for $B=S^7$, which might suggest that some of the theories in \cite{Benishti:2009ky} cannot be interpreted as M2-brane theories.} 
Nevertheless, it would be interesting to study wether in the cases where, after considering the higher $\mathbb{Z}_k$ orbifold, a $\mathbb{C}^2$ is present in the moduli space a M5-spike with $S^3$ profile similar to the one we have found is present.

\vspace{.5cm}

\Large \centerline{\bf Acknowledgements} \vskip 5mm \noindent \normalsize

I am grateful to N.Benishti, D.Berenstein, A. Hanany, Y. Hui-He, J. Sparks for many insightful conversations about different aspects of 3d SCFT's. I would also like to thank Y. Lozano and S. Ramgoolam. Finally, I am grateful to M. Benna and F. Benini for collaborations at the initial stages of this project. I am also grateful to the referee for his/her careful reading of the manuscript. The author acknowledges financial support from the European Commission through Marie Curie OIF grant contract no. MOIF-CT- 2006-38381.

\begin{appendix}
\section{The fuzzy $S^3$ \textit{a la} Castelino, Lee \& Taylor}

Let us start by considering the $SO(5)$ Clifford algebra

\begin{equation}
\{\gamma_{\mu}, \,\gamma_{\nu}\}=2\, \delta_{\mu\nu}\ ,\qquad \mu,\, \nu=1\cdots 5
\end{equation}

These $\gamma_{\mu}$ matrices act on a 4-dimensional space $V$. One can construct the n-fold symmetrized tensor product space $\mathbb{V}=Sym^n\, V$. We can construct the following endomorphisms of $\mathbb{V}$

\begin{equation}
G_{\mu}=\Big(\gamma_{\mu}\otimes \unity\otimes \cdots\otimes \unity+\unity\otimes \gamma_{\mu}\otimes \unity\cdots \otimes \unity+\cdots\Big)_{Sym}
\end{equation}

By defining $G_{\mu\nu}=\frac{1}{2}\,[G_{\mu},\,G_{\nu}]$ one can check that

\begin{equation}
[G_{\mu\nu},\, G_{\alpha\beta}]=2\, \Big(\delta_{\nu\alpha}\, G_{\mu\beta}+\delta_{\mu\beta}\, G_{\nu\alpha}-\delta_{\mu\alpha}\, G_{\nu\beta}-\delta_{\nu\beta}\, G_{\alpha\mu}\Big)
\end{equation}

and thus are the generators of $SO(5)$. Furthermore

\begin{equation}
[G_{\mu\nu},\,G_{\alpha}]=2\,\Big(\delta_{\nu\alpha}\, G_{\mu}-\delta_{\mu\alpha}\, G_{\nu}\Big)
\end{equation}

On the other hand, $SO(4)$ can be easily embedded into $SO(5)$ if we just remove, say, the $G_{5}$ matrix and perform the same construction for the remaining four matrices. If $i=1,2,3,4$, then we can easily obtain the $SO(4)$ algebra 

\begin{equation}
[G_{ij},\, G_{kl}]=2\, \Big(\delta_{jk}\, G_{il}+\delta_{il}\, G_{jk}-\delta_{ik}\, G_{jl}-\delta_{jl}\, G_{ki}\Big)
\end{equation}

Also, one can see that

\begin{equation}
[G_{ij},\,G_{k}]=2\,\Big(\delta_{jk}\, G_{i}-\delta_{ik}\, G_{j}\Big)
\end{equation}

Since $SO(4)\sim SU(2)\times SU(2)$, the irreps are labeled by a pair of spins $(j_l,j_r)$. Then, the fundamental representation of $SO(5)$ decomposes under $SO(4)$ into $(1/2,0)$ and $(0,1/2)$, that is, $V=V^++V^-$. Thus, the symmetric tensor representations of $SO(5)$ decompose under $SO(4)$ as

\begin{equation}
\mathbb{V}=Sym(V^n)=\oplus_{q=0}^n\, V_{(q,\, n-q)}
\end{equation}

where $V_{(q,\,n-q)}$ stands for a subspace with $q$ factors of $V^+$ and $n-q$ factors of $V^-$. Since each $\gamma_i$ matrix is a map from $V^+$ into $V^-$ and vice versa, the $G_i$ matrices are maps between different irreps in the sum above. 

Denoting  $G_k^2=\sum_i\, G_iG_i$, one can easily show that

\begin{equation}
[G_{ij},\, G_k^2]=0
\end{equation}

Thus, thanks to Schur's lemma, $G_k^2$ is proportional to the identity in each irrep, that is, in each $V_{(q,\,n-q)}$.It is then possible to show that

\begin{equation}
\sum_{i=1}^4\, G_i^{(q)}\, G_i^{(q)}=(4\,n+2\,q\,n-2\,q^2)\, \unity_{(q)}
\end{equation}

Here the $(q)$ stands for the restriction to each one of the spaces $V_{(q,\,n-q)}$ labelled by $q$ in the sum above. It is now easy to guess what the minimal choice (leading to the Guralnik-Ramgoolam $S^3$) is: take the minimal group of $q$'s such that the $G_i$ close and such that their corresponding $G_i^2$ has the same proportionality constant with the identity. It is clear that the minimal number of irreps is 2, so if we consider $q$ and $q+1$ we can fix $q$ as

\begin{equation}
(4\,n+2\,q\,n-2\,q^2)=(4\,n+2\,(q+1)\,n-2\,(q+1)^2)\, \Rightarrow q=\frac{n-1}{2}
\end{equation}

which in turn requires odd $n$. Furthermore, then the two spaces we are signaling in the sum are

\begin{equation}
\mathcal{R}_n=\mathcal{R}_n^++\mathcal{R}_n^-\qquad \mathcal{R}_n^+=V_{\big(\frac{n+1}{2}\,\frac{n-1}{2}\big)} ,\quad\mathcal{R}_n^+=V_{\big(\frac{n-1}{2}\,\frac{n+1}{2}\big)}
\end{equation}

In our case, we cannot restrict to this minimal case, and instead we need to consider all the $q$'s. In particular, this implies that $\vec{G}^2\ne \unity$. Nonetheless, since for each $q$ $G_k^{(q)\, 2}=\rho_q^2\,\unity$, it is natural to assume that each such subspace corresponds to an $S^3$ with radius

\begin{equation}
\rho_q^2=(4\,n+2\,q\,n-2\,q^2)
\end{equation}

Finally, it turns out that the dimension of each representation is

\begin{equation}
{\rm Tr}\,\unity_{(q)}=(n-q+1)\,(q+1)
\end{equation}

\end{appendix}


\begin{thebibliography}{99}
\bibitem{Bagger:2007jr}
  J.~Bagger and N.~Lambert,
  Phys.\ Rev.\  D {\bf 77}, 065008 (2008)
  [arXiv:0711.0955 [hep-th]].

\bibitem{Gustavsson:2007vu}
  A.~Gustavsson,
  Nucl.\ Phys.\  B {\bf 811} (2009) 66
  [arXiv:0709.1260 [hep-th]].
  
\bibitem{Aharony:2008ug}
  O.~Aharony, O.~Bergman, D.~L.~Jafferis and J.~Maldacena,
  JHEP {\bf 0810} (2008) 091
  [arXiv:0806.1218 [hep-th]].
  
\bibitem{Schwarz:2004yj}
  J.~H.~Schwarz,
  JHEP {\bf 0411} (2004) 078
  [arXiv:hep-th/0411077].
  
\bibitem{Aharony:2008gk}
  O.~Aharony, O.~Bergman and D.~L.~Jafferis,
  ``Fractional M2-branes,''
  JHEP {\bf 0811}, 043 (2008)
  [arXiv:0807.4924 [hep-th]].
  
\bibitem{Franco:2008um}
  S.~Franco, A.~Hanany, J.~Park and D.~Rodriguez-Gomez,
  JHEP {\bf 0812} (2008) 110
  [arXiv:0809.3237 [hep-th]].
  
\bibitem{Franco:2009sp}
  S.~Franco, I.~R.~Klebanov and D.~Rodriguez-Gomez,
  JHEP {\bf 0908} (2009) 033
  [arXiv:0903.3231 [hep-th]].
  
\bibitem{Aganagic:2009zk}
  M.~Aganagic,
  arXiv:0905.3415 [hep-th].
  
\bibitem{Davey:2009sr}
  J.~Davey, A.~Hanany, N.~Mekareeya and G.~Torri,
  JHEP {\bf 0906}, 025 (2009)
  [arXiv:0903.3234 [hep-th]].
  
\bibitem{Jafferis:2009th}
  D.~L.~Jafferis,
  arXiv:0911.4324 [hep-th].
  
\bibitem{Benini:2009qs}
  F.~Benini, C.~Closset and S.~Cremonesi,
  arXiv:0911.4127 [hep-th].
  
\bibitem{Fabbri:1999hw}
  D.~Fabbri, P.~Fre', L.~Gualtieri, C.~Reina, A.~Tomasiello, A.~Zaffaroni and A.~Zampa,
  Nucl.\ Phys.\  B {\bf 577} (2000) 547
  [arXiv:hep-th/9907219].
  
\bibitem{Martelli:2008si}
  D.~Martelli and J.~Sparks,
  Phys.\ Rev.\  D {\bf 78} (2008) 126005
  [arXiv:0808.0912 [hep-th]].
  
\bibitem{Hanany:2008fj}
  A.~Hanany, D.~Vegh and A.~Zaffaroni,
  JHEP {\bf 0903} (2009) 012
  [arXiv:0809.1440 [hep-th]].
  
\bibitem{Gaiotto:2007qi}
  D.~Gaiotto and X.~Yin,
  JHEP {\bf 0708} (2007) 056
  [arXiv:0704.3740 [hep-th]].
  
\bibitem{Barnes:2005bm}
  E.~Barnes, E.~Gorbatov, K.~A.~Intriligator, M.~Sudano and J.~Wright,
  Nucl.\ Phys.\  B {\bf 730} (2005) 210
  [arXiv:hep-th/0507137].
  
\bibitem{Intriligator:2003jj}
  K.~A.~Intriligator and B.~Wecht,
  Nucl.\ Phys.\  B {\bf 667} (2003) 183
  [arXiv:hep-th/0304128].
  
\bibitem{Benna:2008zy}
  M.~Benna, I.~Klebanov, T.~Klose and M.~Smedback,
  JHEP {\bf 0809}, 072 (2008)
  [arXiv:0806.1519 [hep-th]].
  
\bibitem{Choi:2008za}
  J.~Choi, S.~Lee and J.~Song,
  JHEP {\bf 0903} (2009) 099
  [arXiv:0811.2855 [hep-th]].
  
\bibitem{Basu:2004ed}
  A.~Basu and J.~A.~Harvey,
  Nucl.\ Phys.\  B {\bf 713} (2005) 136
  [arXiv:hep-th/0412310].
  
\bibitem{Terashima:2008sy}
  S.~Terashima,
  JHEP {\bf 0808}, 080 (2008)
  [arXiv:0807.0197 [hep-th]].
  
\bibitem{Gomis:2008vc}
  J.~Gomis, D.~Rodriguez-Gomez, M.~Van Raamsdonk and H.~Verlinde,
  JHEP {\bf 0809}, 113 (2008)
  [arXiv:0807.1074 [hep-th]].
  
\bibitem{Hanaki:2008cu}
  K.~Hanaki and H.~Lin,
  JHEP {\bf 0809} (2008) 067
  [arXiv:0807.2074 [hep-th]].
  
\bibitem{Nastase:2009ny}
  H.~Nastase, C.~Papageorgakis and S.~Ramgoolam,
  JHEP {\bf 0905} (2009) 123
  [arXiv:0903.3966 [hep-th]].
  
\bibitem{Jafferis:2008qz}
  D.~L.~Jafferis and A.~Tomasiello,
  JHEP {\bf 0810}, 101 (2008)
  [arXiv:0808.0864 [hep-th]].
  
\bibitem{Hanany:2008cd}
  A.~Hanany and A.~Zaffaroni,
  JHEP {\bf 0810}, 111 (2008)
  [arXiv:0808.1244 [hep-th]].
  
\bibitem{Berenstein:2002ge}
  D.~Berenstein,
  JHEP {\bf 0204} (2002) 052
  [arXiv:hep-th/0201093].
  
\bibitem{Berenstein:2009ay}
  D.~Berenstein and M.~Romo,
  arXiv:0909.2856 [hep-th].
  
\bibitem{Forcella:2008bb}
  D.~Forcella, A.~Hanany, Y.~H.~He and A.~Zaffaroni,
  JHEP {\bf 0808}, 012 (2008)
  [arXiv:0801.1585 [hep-th]].
  
  \bibitem{AlgebraicGeometry}
  K.E.Smith, L.Kahanpaa, P.Kekalainen and W.Traves, \textit{An invitation to Algebraic Geometry}, Universitext, Springer, 2000.
  
\bibitem{Benvenuti:2006qr}
  S.~Benvenuti, B.~Feng, A.~Hanany and Y.~H.~He,
  JHEP {\bf 0711}, 050 (2007)
  [arXiv:hep-th/0608050].
  
\bibitem{Feng:2007ur}
  B.~Feng, A.~Hanany and Y.~H.~He,
  JHEP {\bf 0703}, 090 (2007)
  [arXiv:hep-th/0701063].
  

\bibitem{Benishti:2009ky}
  N.~Benishti, Y.~H.~He and J.~Sparks,
  arXiv:0909.4557 [hep-th].
  
\bibitem{Ahn:2009bq}
  C.~Ahn and K.~Woo,
  arXiv:0908.2546 [hep-th].
  
  
\bibitem{Guralnik:2000pb}
  Z.~Guralnik and S.~Ramgoolam,
  JHEP {\bf 0102} (2001) 032
  [arXiv:hep-th/0101001].
  
\bibitem{Ramgoolam:2001zx}
  S.~Ramgoolam,
  Nucl.\ Phys.\  B {\bf 610} (2001) 461
  [arXiv:hep-th/0105006].

\bibitem{Ramgoolam:2002wb}
  S.~Ramgoolam,
  JHEP {\bf 0210} (2002) 064
  [arXiv:hep-th/0207111].
  
\bibitem{Castelino:1997rv}
  J.~Castelino, S.~Lee and W.~Taylor,
  Nucl.\ Phys.\  B {\bf 526} (1998) 334
  [arXiv:hep-th/9712105].
  
\bibitem{Klebanov:2009kp}
  I.~R.~Klebanov, S.~S.~Pufu and F.~D.~Rocha,
  JHEP {\bf 0906} (2009) 019
  [arXiv:0904.1009 [hep-th]].
  
\bibitem{Davey:2009bp}
  J.~Davey, A.~Hanany and J.~Pasukonis,
  arXiv:0909.2868 [hep-th].
  
\bibitem{Hewlett:2009bx}
  J.~Hewlett and Y.~H.~He,
  arXiv:0909.2879 [hep-th].
  
\bibitem{Davey:2009qx}
  J.~Davey, A.~Hanany, N.~Mekareeya and G.~Torri,
  arXiv:0908.4033 [hep-th].
  
\bibitem{Imamura:2009ph}
  Y.~Imamura,
  arXiv:0903.3095 [hep-th].
  
\bibitem{Taki:2009wf}
  M.~Taki,
  arXiv:0910.0370 [hep-th].
  
\end{thebibliography}
\end{document}